\documentclass[aps,prb,twocolumn,amsmath,amssymb,nofootinbib,superscriptaddress,floatfix]{revtex4-1}
\usepackage{amssymb}

\usepackage{hyperref}
\usepackage{color}
\usepackage{graphicx}
\usepackage{srcltx}

\begin{document}
\title{
    Symmetry-protected topological phases and
    orbifolds: Generalized Laughlin's argument
}

\date{\today}

\author{Olabode Mayodele Sule}
\affiliation{
Department of Physics,
University of Illinois at Urbana-Champaign,
1110 West Green St, Urbana,
Illinois 61801, USA
            }
\author{Xiao Chen}
\affiliation{
Department of Physics,
University of Illinois at Urbana-Champaign,
1110 West Green St, Urbana,
Illinois 61801, USA
            }

\author{Shinsei Ryu}
\affiliation{
Department of Physics,
University of Illinois at Urbana-Champaign,
1110 West Green St, Urbana,
Illinois 61801, USA
            }

\begin{abstract}
We consider non-chiral symmetry-protected topological phases of matter 
in two spatial dimensions protected by a discrete symmetry  
such as $\mathbb{Z}_K$ or $\mathbb Z_K \times \mathbb Z_K $ symmetry. 
We argue that modular invariance/noninvariance
of the partition function of the one-dimensional edge theory can be used to diagnose whether, by adding a
suitable potential, the edge theory can be gapped or not without breaking the symmetry. By taking bosonic phases
described by Chern-Simons
K-matrix theories and fermionic phases relevant to topological superconductors
as an example, we demonstrate explicitly that when the modular invariance is achieved, we can construct an interaction potential that is consistent with the symmetry and can completely gap out the edge.
\end{abstract}


\maketitle

\section{Introduction} 

Topological phases are highly quantum states of matter that have no classical analog.
The earliest realization of a genuine topological phase
is the quantum Hall effect (QHE). 
Since the discovery of the QHE,  
many exotic properties of topological phases,
their realizations beyond the QHE in two-dimensional electron gas
(e.g., in the context of spin liquids),
and 
their potential use in quantum computation, 
etc.,  
have actively been studied both 
theoretically and experimentally. 
\cite{Wenbook,reviewQHE, NayakReview}

Although symmetry does not play much role in the QHE,
an interplay between topology and symmetry can, and does,
play an important role in the physics of topological phases. A
symmetry-protected topological (SPT) phase is a short-ranged
entangled state of matter with trivial topological order but with
symmetry. That is, in the absence of symmetry all such phases
can be connected adiabatically to one another. With symmetry,
however, they are distinct phases that cannot be adiabatically
connected to one another without breaking the symmetry.
Examples of SPTs include, for example, the Haldane phase in
one-dimensional quantum spin chains, the quantum spin Hall
effect, and the three-dimensional time-reversal-symmetric
topological insulator and superconductors.
\cite{
reviewTIa,
reviewTIb,
KaneMele,
Bernevig05,
Bernevig06,
Moore06,
Roy3d,
Fu06_3Da,
Fu06_3Db,
qilong,
Schnyder2008,
SRFLnewJphys, 
Kitaev2009}

In one spatial dimension (d
=
1), (topological) properties
of gapped phases can be conveniently studied in terms of
the tensor-network (matrix-product state) representation of the
ground-state wave functions.
\cite{Pollmann et al, ChenWenGu2011,
Schuch et al, 
ChenWenGu2011b, FidkowskiKitaev2010, FidkowskiKitaev2011, Turner2011, Gu09}
By utilizing such representation of the ground-state wave functions, it has been shown
that all SPT phases in
$d=1$ with a given symmetry group
G can be fully classified. \cite{Pollmann et al, ChenWenGu2011,
Schuch et al, 
ChenWenGu2011b}
In higher dimensions
$d>1$, a
large class of ground-state wave functions of SPT phases can
be constructed using the tensor-network method.\cite{Chen2011, Gu12}
For SPT
phases that can be realized by combining Abelian topological
phases in two dimensions, Chern-Simons
K-matrix theories
have been adopted to construct, characterize, and classify
various SPT phases. \cite{LevinStern2009, Neupert2011, LuVishwanath2012, LuVishwanath2013, Wang, Kapustin} Another theme involves using the
braiding statistics of quasiparticle excitations in the bulk to
classify the SPT phases.\cite{Levin 2013, HanWan2013, ChengGu2013, LevinGu2012, LevinGu}

The purpose of this paper is to discuss SPT phases in 2+1 dimensions[(2+1)D]
that are accompanied by stable edge states,
following the approach proposed in Ref.\ \onlinecite{RyuZhang2012}. 
Suppose we have a (1+1) dimensional conformal field theory (CFT),
that may be realized as an edge state of a (topological) bulk theory. 
The CFT can be either chiral or non-chiral, but our main focus will
be on non-chiral edge theories. 
Let us also assume the CFT is invariant under some global on-site symmetry group
$G$. 
We then ask if the gapless nature of the CFT is stable or not
against perturbations  
once the symmetry $G$ is enforced.
It the CFT cannot be gapped without breaking the symmetry $G$, 
it cannot exist on its own as an isolated (1+1)-dimensional system,
but it must be realized as an edge theory of a bulk SPT protected by
symmetry group $G$. 

The strategy suggested in Ref.\ \onlinecite{RyuZhang2012}
consists of the following two ingredients. 
The first ingredient is the strict enforcement of the symmetry
by a projection operation on the CFT. 
(A similar projection or gauging procedure 
is also employed in Ref.\ \cite{LevinGu2012} in a related context -- 
see below.)
The second ingredient is the
modular (non) invariance of the projected edge CFTs.
This can be thought of as a geometrical (or: gravitational)
generalization of Laughlin's thought experiment. 
\cite{Laughlin1981}
In the Laughlin argument, 
an insertion of an integer multiple of 
flux quanta in a quantum Hall system on the cylinder geometry
(say the ends of a cylinder) results in a transport of charge excitations 
from one edge to another,
i.e., Thouless pumping.
This leads to a non-conservation (quantum anomaly)
of particle number 
within the single edge theory 
and can thus be viewed as a signature of the presence of a non-trivial bulk. 
This idea can be generalized to include modular transformations 
which are large coordinate transformations in CFTs.

Within our approach, 
the characterization and classification of SPT phases can be viewed as 
those
of (non-chiral) CFTs with symmetry projection. 
Such CFTs are obtained from a projection by
a (discrete) group and are usually called orbifold CFTs.
\cite{orbifolds1, orbifolds2, footnote}
It is often the case that the symmetry-projected 
theory can be made modular invariant as far as the symmetry group
acts on holomorphic and anti-holomorphic sectors of the CFT
in a symmetric fashion. 
Once the symmetry group acts on the holomorphic and anti-holomorphic
sectors in an assymetric way (e.g., as an extreme example, $G$ can act
solely on the holomorphic sector but not on the anti-holomorphic sector)
the modular invariance is no longer guaranteed.
Such orbifold is called an asymmetric orbifold. 
\cite{assymetric} 
Our approach 
may then be summarized as 
``(2+1)D SPT phases = asymmetric orbifolds'', 
as a slogan.  
\cite{footnote}

The structure of the paper and the main results are
summarized as follows. In Sec. II
we start our discussion by
giving a more general discussion of modular invariance and
gauging symmetries in SPT phases.

In Sec.III, we discuss bosonic SPT phases described by
Chern-Simons K-matrix theories in the bulk with
$\mathbb Z_K \times \mathbb Z_K$ symmetry. With $N_f$
copies (flavors) of the edge theories, we
find that modular invariance can be achieved when there are
$N_f=0 $ mod $K$ flavors. This suggests the phases are trivial if and only
if (iff) $N_f=0$ mod $K$. When $N_f=1$ and for a particular $\mathbb Z_
K$ subgroup of $\mathbb Z_K \times \mathbb Z_K$
we also reproduce within the modular
(non)invariance approach some of the results based on a
K-matrix formulation used in Ref.\ \onlinecite{LuVishwanath2012}.

Similarly, in Sec. IV, we study fermionic phases with $\mathbb Z_K \times \mathbb Z_K$ symmetry. We find that modular invariance can be achieved when $N_f = 2K$ (mod $2K$). Here $N_f$ refers to the number of flavors of complex Dirac fermions. This generalizes the earlier
results in Refs. \onlinecite{RyuZhang2012} and \onlinecite{LiangQi}. (See Ref. \onlinecite{YaoRyu2012} for a similar model.) In
all cases considered we analyze the stability of the edge states
to gapping when interactions are considered and explicitly
construct ($\mathbb Z_K \times \mathbb Z_K$)-invariant potentials that can completely
gap out the edge states whenever the phase is trivial. Finally,
we conclude in Sec. V with a summary of our results and an outlook for future applications.

\section{Modular invariance and gauging symmetry in SPT phases}

\paragraph{Laughlin's thought experiment}
We start by giving a brief overview of 
our approach to (2+1) dimensional SPT phases,
which may be viewed as an gravitational 
analog of Laughlin's thought experiment. 
\cite{Laughlin1981} 

Let us first recall some key steps in Laughlin's thought experiment. 
Consider a quantum Hall system on a finite cylinder 
with two edges, I and II. 
If magnetic flux $\Phi$ is threaded adiabatically
through the cylindrical hole, as the flux is increased (adiabatically), starting from, say, zero flux, 
the Hamiltonian $H(\Phi)$ of the system is in general not invariant under the flux insertion. 
However, after an increase by an integral multiple of the flux quantum $\Phi_0$, the Hamiltonian comes back to itself. That is,  
$H(\Phi + n \Phi_0) = H(\Phi), 
$
for any integer $n$. 
This is a symmetry of the system but cannot be achieved 
by successive applications of infinitesimal gauge transformations.

In the process of increasing the flux,
an integer multiple of electric charges actually 
get pumped from one edge to the other
due to the QHE. 
One can analyze this from the point of view of the partition function 
for the excitation spectrum of the system. 
Since the transformation is adiabatic and the bulk spectrum 
is gapped it is enough to focus on the partition function 
of the edge excitations which have a gapless spectrum. 
This is given by
\begin{equation}
 Z(\Phi) = \sum_{a, b} N_{a b} \chi^{\mathrm{I}}_a (\Phi) 
 \chi^{\mathrm{II}}_b(\Phi), 
\end{equation}
where $\chi_a^{\mathrm{I}, \mathrm{II}}$ is a chiral contribution to the partition function for each edge and $N_{ab}$ are constants.
Under a large gauge transformation in general,
the chiral parts of the partition function is not invariant, 
$\chi_a(\Phi + n \Phi_0) \neq \chi_a(\Phi)$,
while the total partition function is invariant. 
This is to be expected for a system with non-trivial bulk and signals the non-conservation of charge. 
\cite{Cappelli}
In the case of the quantum spin Hall effect, 
a similar flux threading argument
\cite{LevinStern2009, qshe_flux}
can be applied to show that 
a flux change by $\Phi_0 /2$ pumps fermion number parity 
and leads to spin charge separation.

In the cases we will be interested in, since the total charge may not necessarily be conserved,
we look for an alternative to the flux $\Phi$. As advocated in the Introduction, a natural thing to do is to simply replace $\Phi$ by  the modular parameter of the torus $\tau$. 

\paragraph{CFT and modular invariance}
The effective continuum theory describing 
the gapless edge state of a bulk (topological) phase is a CFT. 
A CFT is a field theory defined on a Riemann surface,
i.e., a surface with a choice of complex structure (complex coordinates). 
Two complex structures on a surface are equivalent if there 
is a biholomorphic map between them. 
Classically a CFT is invariant under a transformation 
between equivalent complex structures. 

In physical terms CFTs are usually called scale invariant theories, 
the scale invariance here can be thought of as a rescaling of the metric. On a 2-dimensional surface there is a one to one correspondence beween the space of conformal classes of metrics (with metrics related by a rescaling considered equivalent) and the space of complex structures. Thus a rescaling of the metric can be thought of as a small gauge transformation which via the correspondence is simply a holomorphic(complex-analytic) change of coordinates.
After quantization the rescaling invariance of CFTs suffer an anomaly, 
the trace anomaly proportional to the central charge. 

A more general type of gauge transformation is the so called 
large gauge transformation that cannot be continuously deformed 
to the identity.  
On the simplest lower genus surfaces like the two-sphere and the plane 
there are no large conformal gauge transformations.  
If one considers 
the torus one finds that the set of all complex tori 
(tori with complex structure) are in correspondence with points 
in the upper half plane. 
A generic point $\tau$ in the upper half plane represents a torus.
It turns out that under a transformation
\begin{equation}\label{mod}
\tau \rightarrow \frac{a\tau + b}{c\tau +d}
\end{equation}
with $a, b, c, d \in \mathbb Z$ and $ad-bc=1$ one obtains a torus that is equivalent in the sense defined above. These are the large gauge (holomorphic coordinate) transformations on a torus and are referred to as modular transformations. 

Modular invariance is the statement of invariance under the transformation (\ref{mod}) 
and it is usually taken as a constraint on any CFT, if it is derived as the continuum limit of a two-dimensional lattice statistical mechanical system, or a (1+1)-dimensional lattce quantum system.\cite{Cardy1986}
Modular invariance can be thought of as a basic constraint that ensures that a CFT is invariant under large gauge transformations at higher genus. 
Amongst other things it imposes some constraints on the operator content of any CFT,
This is the case because a CFT on higher genus can always be constructed from CFT on lower genus by sewing and cutting surfaces,
\cite{Polchinski98} and the large gauge transformations at higher genus are related to those at lower genus. 
Modular transformations are generated by the so called Dehn twists $ \tau \rightarrow \tau +1$ and $\tau \rightarrow -1/\tau$. 
In terms of the usual representation of a torus as a doughnut the second transformation exchanges 
the two nontrivial cycles and can be thought of as an exchange of space and time.

For our application to the edge states, we take our (2+1)-
dimensional bulk systems to exist on a cylinder as before.
This means that each edge is topologically $S^1$
and as far as thermal physics is concerned, the Euclidean field theory for
the gapless edge states is a CFT on a geometry which is
topologically a torus $S^1 \times S^1$. There is an obvious analogy
between the large gauge transformations in the original
Laughlin’s argument and the large coordinate transformation,
i.e., the modular transformations. We are thus led to claim
that inability to achieve modular invariance is a signal of
a nontrivial bulk: Focusing on a single-edge theory (edge
I or II in the above discussion), if one cannot construct a
modular-invariant partition function within the edge theory,
the edge theory cannot be gapped. In other words, such a
theory cannot be realized as an isolated (1+
1)-dimensional system and must be realized as an edge theory of a nontrivial
(2+1)-dimensional bulk state. Conversely, if one can achieve
modular invariance, the edge theory can be gapped.

\paragraph{symmetry projection}

The argument given above so far has not referred to the role
played by symmetries. To discuss SPT phases, we now discuss
the interplay between symmetry conditions in SPT phases and
the modular (non)invariance.

Quite often, a non-chiral CFT can be made modular invariant,
once holomorphic and antiholomorphic sectors are properly combined. 
The non-chiral edge theory of a SPT phase is no exception
if symmetry conditions are not enforced --
without enforcing symmetries, a SPT phase is adiabatically deformable
to a trivial state. 
There may however be a conflict between 
modular invariance 
and symmetry conditions, as the latter may forbid
particular ways of combining the left- and right-moving parts of the CFT.

Our strategy to diagnose and characterize
a SPT phase is then, 
to impose the symmetry conditions strictly and ask
if the system is invariant under modular transformations or not.  
More specifically, we take only a single-edge theory (edge
I or II in the above notation) and ask if it can be made
modular invariant. (If we consider the two separate edges of
the modular anomalous theories, they can be combined in a
modular-invariant way.) We achieve the strict enforcement of
the symmetry by considering a projection of the CFT by the
symmetry group
G. We now take a close look at the projection
procedure.

Let us consider a 1+1 D edge theory with field $\phi$ that has some global symmetry. 
We work on the torus and given a finite subgroup $G$ of the center of the global symmetry group,
we would like to ``gauge'' $G$. 
In other words, we would like to consider a field theory with the same Hamiltonian as before but with Hilbert space restricted to the $G$ invariant subspace of the original 
(this is valid because the Hamiltonian takes $G$ invariant states to $G$ invariant states). 
In the Hilbert space formalism, the restriction to $G$ invariant states 
is the same thing as a projection, i.e the partition function in the $G$ invariant sector is 
\begin{equation}
Z= \mathrm{Tr}\, P e^{-tH}, 
\end{equation}
where $P = |G|^{-1} \sum_{g\in G} \hat g$ is a projection operator satisfying $P^2 = P$, as is easily verified. In the path integral formalism this has the interpretation as a sum over all fields twisted  in the time direction of the torus i.e 
\begin{equation}\label{generalpartition}
Z = \sum_{g\in G} \int \mathcal{D} [\phi^g]
e^{-S[\phi]}, 
\end{equation}
where the subscript $g$ indicates that the path integral is over fields satisfying 

\begin{equation}\label{boundary}
\phi(x + 2\pi \tau_1, t+ 2\pi \tau_2) = g\phi(x, t)g^{-1}.  
\end{equation}
This is the meaning of twisting in the time direction. (Here in addition to the translation in the time direction by $2\pi\tau_2$, a finite shift $2\pi\tau_1$ in the spatial direction is also added. This is necessary to discuss a CFT defined on a general torus specified by the modular parameter $\tau$. These two parameters are combined into the modular parameter $\tau$, $\tau = \tau_1 + i \tau_2$.)

Since the modular invariance is a constraint for a gapless edge theory 
one needs to ensure that that the partition function (\ref{generalpartition}) is modular invariant. 
Considering that S modular transformations interchange space and time 
it is clear that to achieve modular invariance one would have to 
explicitly include sectors in the path integral twisted 
in the spatial direction. 
[On the other hand, 
the T modular invariance essentially imposes a constraint on the spectrum of dimensions (hence on operators) in the ``gauged'' (projected) theory.]
One can propose that a suitable partition function for the gauged theory is given by 
\begin{equation}
Z = \sum_{(h,g)\in G^2} \int \mathcal{D} [\phi^h_g]
e^{-S[\phi]}, 
\end{equation}
where now the subscript $(h,g)$ indicates that the path integral is restricted to fields twisted in the spatial and time directions by $h$ and $g$ respectively: In addition to the boundary condition (\ref{boundary}), the field also obeys

\begin{equation}
\phi(x + 2\pi, t) = h\phi(x, t)h^{-1}.  
\end{equation}

In general,
if the space of all possible field configurations in
the path-integral formulation of a quantum field theory is disconnected, 
the partition function is given as a sum over all disconnected sectors, 
$Z=
\sum_n
\int \mathcal{D} [\phi_n]  
\epsilon_n
\exp(-S[\phi_n]),
$
where $n$ labels disconnected components of the field configurations.  
There is no {\it a priori }way
to fix the relative amplitude $\epsilon_n$. 
When the relative amplitudes between different sectors
are complex,
this is the physics of topological terms 
(such as the topological term in the non-linear sigma models)
in general. 
We are thus lead to consider  
\begin{equation}
 Z = \sum_{(h,g)\in G^2} \epsilon^h_g\int \mathcal{D} [\phi^h_g]
e^{-S[\phi]}, 
\end{equation}
where the $\epsilon^h_g$ are phases that assign different weights to different topological sectors.
The phase factors $\epsilon_g^h$, which are called
discrete torsion, can be viewed as arising from 
projection onto different sectors,
or, alternatively, different
ways to assign the quantum number for the ground state in each sector.


\paragraph{duality between SPT phases and 
topological phases}
To summarize,
we consider the partition function orbifolded (twisted) by the symmetry group of the problem,
and ask if the partition function is modular invariant or not. 
We view this as a generalization of Laughlin's flux threading
 argument. 
 It is a generalization in the following two sense:
 (i) it is a geometrical generalization
 (ii) symmetry plays an important role. 
The idea of orbifolding the edge CFT of SPT phases
can be viewed in an alternative way, 
as suggested in Ref \onlinecite{LevinGu2012}. (See also Ref \onlinecite{Levin 2013}.)
These authors proposed that if the global symmetry of an SPT phase with symmetry group G is promoted to a gauge symmetry, the resulting phase is a topologically ordered phase with a particular pattern of fractional statistics.

To make a connection with our approach,
we first note that the gauging and symmetry projection
have a similar effect in that we focus on a gauge singlet sector
of the theory
[although the gauging means in general 
imposing singlet condition locally (e.g., at each of a lattice),
while projection is enforced only globally].\cite{footnote2} Next, for Abelian theories (multicomponent
chiral/nonchiral bosons compactified on a lattice) a self-dual
condition together with an even-lattice condition guarantees
that modular invariance is achieved. The same condition can
be derived from the argument based on fractional statistics as shown in Ref. \onlinecite{Levin 2013}.

\section{Abelian bosonic symmetry-protected topological phases} 

In this section, 
we discuss bosonic 
SPT phases that can be described in terms of 
abelian Chern-Simons theories. 
At the boundary (edge), they support excitations described by
a (1+1) dimensional (non-chiral) boson theory. 
Since it is non-chiral, the edge theory is 
unstable unless certain discrete symmetry condition $G$ is imposed. 

One can view  this discrete symmetry as embedded in the $U(1)^N\times U(1)^N$
continuous symmetry of the Abelian Chern-Simons theory.
Within this setting, the gauging can be achieved by the Higgs mechanism. 
The idea is that by a choice of suitable Higgs fields potentials 
and couplings one can break the symmetry to the subgroup $G$. 
One can then study the SPT phase with symmetry group $G$ as 
a symmetry broken phase of a spontaneously broken gauge symmetry.
It is essentially a gauge theory with (discrete) gauge group $G$. 
The bulk (2+1)-dimensional Lagrangian for such a system is given by 
\begin{equation}
 \mathcal{L}
 =
 \frac{K_{IJ}}{4\pi} \epsilon^{\mu\nu\lambda} a^I_{\mu} \partial_\nu a^J_\lambda 
 + q^Ij_I^\mu A_\mu + \mathcal {L}_{\mathrm{Higgs}}[\{\psi\}, A], 
\end{equation}
where $K_{IJ}$ is a $2 N\times 2N$ integer valued matrix with 
$|\mathrm{det}\, K_{IJ}| = 1$.
(We shall consider below $N_f$ flavors of such theories in which case $K_{IJ}$ becomes 
a block diagonal $2NN_f \times 2NN_f$ matrix).
The gauge field $A_{\mu}$ is an external probe field which couples minimally 
to the internal current 
$j^\mu = \frac{1}{2\pi} \epsilon^{\mu\nu\lambda} \partial_\nu a_\lambda$ with the corresponding charge $q_I$.
The $\{\psi\}$ are Higgs fields chosen in appropriate representations to break the symmetry to $G$.
The internal fields $a^I_{\mu}$ can be integrated out to obtain 
an effective Chern-Simons Higgs theory for the external field $A_{\mu}$.

We shall simply consider the edge theory for a bosonic state which for a non-chiral state (after turning off the external potential) can be taken without loss of generality to be $K_{IJ} = \sigma_x$ ($\sigma_x$ is the $x$ component of the Pauli matrices), by using $GL(2, \mathbb Z)$ symmetry $K_{IJ} \rightarrow W^T K _{IJ}W$ with 
$|\mathrm{det}\, W| = 1$.\cite{LuVishwanath2012} In this case the edge theory is obtianed using a gauge invariance argument of the Chern-Simons theory to be
\begin{align}\label{bosonedge}
S = \frac{1}{4\pi} 
\int dt dx \left(\partial_t \phi_1 \partial_x \phi_2 + \partial_t \phi_2 \partial_x \phi_1- V_{IJ}\partial_x \phi_I\partial_x\phi_J\right). 
\end{align}
(We start with the case of one flavor $N_f=1$ and we introduce flavors later).

We consider $G = \mathbb Z_K \times \mathbb Z_K$ and consider the symmetry transformation 
for $(k_1,l_1) \in \mathbb Z_K \times \mathbb Z_K$
\begin{align}\label{symmetry}
\phi_1 \rightarrow \phi_1 + \frac{2\pi k_1}{K},
\quad 
\phi_2 \rightarrow \phi_2 + \frac{2\pi l_1}{K}.
\end{align}
We note that the fields have the periodicity $\phi_I = \phi_I + 2\pi$ 
so that this is indeed a $\mathbb Z_K \times \mathbb Z_K$ symmetry transformation. 
Setting $l_2 = l_1 q$ for $q\in \mathbb Z$ reduces 
to a $\mathbb Z_K$ subgroup of $\mathbb Z_K \times \mathbb Z_K$ 
which is one of the cases considered in Ref. \onlinecite{LuVishwanath2012}.
We shall analyze the modular invariance properties of the gauged CFT described by 
the action (\ref{bosonedge}) and symmetry transformation (\ref{symmetry}).

In terms of the chiral modes $\phi_L$ and $\phi_R$ defined by 
\begin{align}
\phi_1 &= \sqrt{\frac{r}{2}}(\phi_L +\phi_R), 
\quad
\phi_2 = \sqrt{\frac{1}{2r}}(\phi_L-\phi_R), 
\end{align}
one can rewrite this as
\begin{align}
S = \frac{1}{4\pi} \int dt dx 
\left[ \partial_t \phi_L  \partial_x \phi_L  - v(\partial_x \phi_L)^2\right.
\nonumber\\
\left.-\partial_t \phi_R \partial_x\phi_R - v(\partial_x \phi_R)^2\right]
\end{align}
where $r := \sqrt{\frac{V_{22}}{V_{11}}}$ and we have assumed $V_{12} + V_{21} = 0$ so that we obtain a non-chiral theory with equal left and right moving velocities $v =2 \sqrt {V_{11} V_{22}}$. 
 Henceforth, we set 
 $v=1$ and the system size $L=2\pi v$, 
 without loss of generality. 
The Hamiltonian and total momentum are given by 
\begin{align}
H &= \frac{1}{4\pi} \int dx  
\left[(\partial_x \phi_L)^2 + (\partial_x \phi_R)^2\right],
\nonumber\\
P&=\frac{1}{4\pi} \int dx 
\left[
 (\partial_x \phi_L)^2 -  (\partial_x \phi_R)^2
\right]. 
\end{align}

The quantization of the theory is pretty standard. 
The equal time canonical commutation relations are given by
\begin{align}\label{commutators} 
[\partial_x \phi_L(x,t), \partial_y \phi_L(y,t)] &= 2\pi i\partial_x \delta(x-y),
\nonumber\\
[\partial_x \phi_R(x,t), \partial_y \phi_R(y,t)]&= -2\pi i\partial_x \delta(x-y). 
\end{align}
From the equations of motion one can deduce the mode expansions
\begin{align}
\phi_L(x,t) &= \phi_{L,0} + p_L (t+x)+ i\sum_{n \neq 0}  \frac{a_n}{n} e^{-i n(t+x)}, 
\nonumber \\
\phi_R(x,t) &= \phi_{R,0} + p_R (t-x)+ i\sum_{n \neq 0}  \frac{b_n}{n} e^{-i n(t-x)}, 
\end{align}
so that Eq.\ (\ref{commutators}) translates into
\begin{align}
[a_n, a_m] = [b_n, b_m] = n \delta_{n+m,0}\nonumber\\
[\phi_{R,L}, p_{R,L}]= i
\end{align}
(We will mostly work in the basis where $p_{R,L}$
are diagonal, i.e.,they can be thought of as $c$
numbers. The allowed values of
the zero mode $p_{L,R}$ will be determined from the boundary
conditions below.)
Therefore $a_n$ and $b_n$ are the oscillator modes of a free boson. The quantum Hamiltonian and total momentum in terms of the oscillators are 
\begin{align}\label{Hamiltonian momentum}
H &= 
\frac{p_L^2}{2} + \sum_{n > 0} a_{-n}a_n-\frac{1}{24}
+
\frac{p_R^2}{2} + \sum_{n > 0} b_{-n}b_n-\frac{1}{24},   
\nonumber\\
P &=\frac{ p_L^2 -p_R^2}{2} + \sum_{n>0} 
\left(a_{-n}a_n -b_{-n}b_n\right). 
\end{align}
The constant term in $H$ is due to the regularization of the sum of zero point energies $\sum_n n  \rightarrow -{1}/{12}$.\\

\subsection{$\mathbb Z_K \times \mathbb Z_K$ symmetry}

We are interested in gauging the ${\mathbb Z}_K\times{\mathbb Z}_K$ symmetry (\ref{symmetry}).
Therefore on a torus of modular parameter $\tau$ we need to consider sectors with twisted spatial periodicities
\begin{align}
\phi_1(x+ 2\pi, t) &= \phi_1(x,t) + 2\pi \Big(n + \frac{k_1}{K}\Big),
\nonumber\\
\phi_2(x+ 2\pi, t) &= \phi_2(x,t) + 2\pi \Big(m + \frac{l_1}{K}\Big). 
\end{align}
where $n,m \in \mathbb Z$. In terms of $\phi_L$ and $\phi_R$, this is 
\begin{align}
\phi_L(x+2\pi, t) &= \phi_L(x,t)+
\frac{2\pi}{\sqrt {2r}}\left[ (n+ \frac{k_1}{K}) + r (m+\frac{k_1}{K})\right], 
\nonumber\\
\phi_R(x+2\pi, t) &= \phi_R(x,t)+
\frac{2\pi}{\sqrt {2r}}\left[(n+ \frac{k_1}{K}) - r (m+\frac{l_1}{K})\right].
\end{align}
This leads to the following quantization condition for the zero internal momentum modes
\begin{align} \label{Zero momentum}
p_L &= \frac{1}{\sqrt {2r}}
\left[
(n+ \frac{k_1}{K}) + r (m+\frac{l_1}{K})\right],
\nonumber\\
p_R &= \frac{1}{\sqrt {2r}}
\left[ -(n+ \frac{k_1}{K}) + r (m+\frac{l_1}{K})
\right].
\end{align}

We would also need an expression for the operator $(\hat k_2, \hat l_2)$ that implements translations by $(k_2, l_2)$ in the Hilbert space. This can be deduced by computing the commutators between the field $\phi_{1,2}$, and the $U(1)$ charges
\begin{align}
Q_1:= \frac{1}{2\pi}\int dy \partial_y \phi_2(y,t) = \sqrt{\frac{1}{2r}}(p_L+p_R),\nonumber\\
Q_2:= \frac{1}{2\pi}\int dy \partial_y \phi_1(y,t) = \sqrt{\frac{r}{2}}(p_L-p_R).
\end{align}
Since $[\phi_1(x,t), Q_1(t)] = [\phi_2(x,t), Q_2(t)] = i, Q_{I=1,2}$ generates translations of $\phi_I$. The desired operator is then given by
\begin{align}
 \label{time twist}
(\hat k_2, 
\hat l_2) 
&= 
\exp
\frac{2\pi i}{K}
\left[  k_2Q_1
+  l_2Q_2
 \right]
\nonumber\\
&= 
\exp
i \left[
 (m+ \frac {l_1}{K}) \frac{2\pi k_2}{K} +  (n+ \frac{k_1}{K})\frac{2\pi l_2 }{K} \right]
\end{align}

\subsubsection{Partition Function}

We are interested in modular properties of the gauged partition function 
\begin{align}
Z = \frac{1}{K^2} \sum_{k_1,k_2,l_1,l_2=0}^{K-1} 
\epsilon^{(k_1,l_1)}_{(k_2, l_2)} 
Z^{(k_1,l_1)}_{(k_2,l_2)}(\tau)
\end{align}
with
\begin{align}
 Z^{(k_1,l_1)}_{(k_2,l_2)} = 
\mathrm{Tr}_{k_1,l_1} 
\left[
 (\hat k_2, \hat l_2) e^{2\pi i P\tau_1-2\pi \tau_2 H}
 \right]
\end{align}
being the partition function in the sector twisted by $(k_1, l_1)$ and $(k_2,l_2)$ in the spatial and time directions, respectively. Using the previous results (\ref{Hamiltonian momentum}, \ref{Zero momentum}, \ref{time twist}) one obtains
\begin{align}
 \label{partition1}
&
Z^{(k_1, l_1)}_{(k_2,l_2)} (\tau)
= |\eta(\tau)|^{-2}
\sum_{(n,m) \in {\mathbb Z}^2}
\nonumber \\
&\quad
\times
\exp
\left\{\frac{2\pi i k_2}{K} \left(m +\frac{l_1}{K}\right) 
+ \frac{2\pi i l_2}{K}\left(n +\frac{k_1}{K}\right)
\right. 
\nonumber \\
&\qquad 
-\pi \tau_2
\left[\frac{1}{r} \left(n+\frac{k_1}{K} \right)^2 + r \left(m +\frac{l_1}{K}\right)^2\right]
\nonumber\\
&
\qquad \left.+ 2\pi i \tau_1 \left(n+ \frac{k_1}{K}\right) \left(m + \frac{l_1}{K}\right) \right\}, 
\end{align}
where $\tau = \tau_1 + i \tau_2$, and $\eta(\tau)$ is the Dedekind eta function which arises as a result of the sum over the modes of the oscillators.

\subsubsection{Large gauge anomaly and modular invariance}

Let us explore the behaviour of Eq. (\ref{partition1})
under the ``large gauge transformations" $k_1 \rightarrow k_1\pm K$ and similarly for $k_2, l_1$ and $l_2$.  Under $k_1 \rightarrow k_1+K$ and $l_1 \rightarrow l_1 +K$ by relabelling $n\rightarrow n+1$ and $m\rightarrow m+1$ respectively it is easy to see that Eq. (\ref{partition1}) is invariant. On the other hand under hand one has the following large gauge anomaly:
\begin{align}\label{anomaly1}
 Z^{(k_1, l_1)}_{(k_2\pm K,l_2)} (\tau)
&= e^{\frac{\pm 2\pi i l_1}{K}} 
Z^{(k_1, l_1)}_{(k_2,l_2)} (\tau),
\nonumber\\
Z^{(k_1, l_1)}_{(k_2,l_2\pm K)} (\tau) 
&= e^{\frac{\pm 2\pi i k_1}{K}} 
Z^{(k_1, l_1)}_{(k_2,l_2)} (\tau).  
\end{align}

Under $T$ modular transformations, $\tau \rightarrow \tau +1$, one finds
\begin{equation}\label{ttransformation1}
 Z^{(k_1, l_1)}_{(k_2,l_2)}(\tau+ 1 ) 
 = e^{-\frac{2\pi i k_1l_1}{K^2}} Z^{(k_1, l_1)}_{(k_1+k_2,l_1+ l_2)}(\tau). 
\end{equation}
To determine the behavior under $S$ modular transformations, it is convenient to use the Poisson resummation formula  $\sum_{n\in \mathbb Z} f(n) = \sum_{n\in \mathbb Z} \int_{-\infty}^\infty f(x) e^{-2\pi i x n} dx$ to  rewrite Eq. (\ref{partition1}) as 
\begin{align}
 &Z^{(k_1, l_1)}_{(k_2,l_2)}(\tau) = 
|\eta(\tau)|^{-2}\sqrt{\frac{r}{\tau_2}} 
 \sum_{(n,m) \in {\mathbb Z}^2} 
\nonumber \\
&\quad \times \exp\left\{
 \frac{2\pi i k_2}{N} \left(m+\frac{l_1}{K}\right) 
+ \frac{2\pi i k_1 n}{K} 
 \right.
\nonumber\\
&\quad \quad
\left. -\frac{\pi r}{\tau_2} \left[|\tau|^2 \left(m+\frac{l_1}{K}\right)^2 + \left(\frac{l_2}{K}-n\right)^2\right]\right.
\nonumber\\
&\quad \quad
\left.+\frac{2\pi r\tau_1}{\tau_2} 
\left(nm + \frac{l_1n}{K} -\frac{l_2m}{K}-\frac{l_1l_2}{K^2}
\right)
\right\}.
\end{align}
Thus, under $S$,
i.e.,
$\tau_1 \rightarrow -\frac{\tau_1}{|\tau|^2}$,
$\tau_2 \rightarrow \frac{\tau_2}{|\tau|^2}$, 
by switching $n \rightarrow -m, m \rightarrow n$ one finds 
\begin{equation}\label{stransformation1}
 Z^{(k_1, l_1)}_{(k_2,l_2)} (-1/\tau) = 
 e^{\frac{2\pi i (l_1k_2+k_1l_2)}{K^2}} 
 Z^{(k_2, l_2)}_{(-k_1,-l_1)} (\tau).
\end{equation}

The total partition function with $N_f$ flavors and $\mathbb Z_K \times \mathbb Z_K$ gauge invariance is given by
\begin{align}
Z (\tau) = \sum_{k_1,l_1,k_2,l_2 = 0}^{K-1} 
\epsilon^{(k_1,l_1)}_{(k_2,l_2)}
\left[Z^{(k_1, l_1)}_{(k_2,l_2)}\right]^{N_f},
\end{align}
with $\epsilon^{(k_1,l_1)}_{(k_2,l_2)}$ having the same meaning as before.
From Eq. (\ref{stransformation1}) one determines the following conditions for $S$ modular invariance
\begin{align} \label{srule1}
&
\epsilon^{(k_1,l_1)}_{(k_2, l_2)}= 
\nonumber \\
&\quad
\begin{cases}
 e^{-\frac{2\pi i N_f(l_1k_2+k_1l_2) }{K^2}} \epsilon^{(k_2, l_2)}_{(-k_1,-l_1)}
\quad \text{if } k_1=l_1=0\\
\\
e^{-2\pi i N_f \left[\frac{(l_1k_2+k_1l_2) }{K^2} -\frac{l_2}{K}\right]} 
\epsilon^{(k_2, l_2)}_{(-k_1+K,-l_1)} 
\quad \text{if } l_1=0 ,k_1 > 0
\\ \\
e^{-2\pi i N_f \left[\frac{(l_1k_2+k_1l_2) }{K^2} -\frac{k_2}{K}\right]} 
\epsilon^{(k_2, l_2)}_{(-k_1,-l_1+K)}  
\quad \text{if } k_1=0, l_1>0\\
\\
 e^{-2\pi i N_f
  \left[\frac{(l_1k_2+k_1l_2) }{K^2} -\frac{k_2+l_2}{K}\right]}
  \epsilon^{(k_2, l_2)}_{(-k_1+K,-l_1+K)} \\ 
 \qquad \text{if } k_1,l_1>0
\end{cases}
\end{align}
while from (\ref{ttransformation1}) one determines the following conditions for T modular invariance:
\begin{align}\label{trule1}
&
\epsilon^{(k_1,l_1)}_{(k_2, l_2)} =
\nonumber \\
&\quad
\begin{cases}
  e^{\frac{2\pi i N_f l_1k_1 }{K^2}} 
  \epsilon^{(k_1, l_1)}_{(k_1+k_2,l_1+l_2)}
  \\
\qquad \text{if } k_1+k_2, l_1+l_2 \leq K-1\\
\\
e^{2\pi i N_f\left[\frac {l_1k_1}{K^2} -\frac{l_1}{K}\right]}
\epsilon^{(k_1, l_1)}_{(k_1+k_2-K,l_1+l_2)}
\\
\qquad \text{if } k_1+k_2 > K-1, l_1+l_2 \leq K-1\\
\\
e^{2\pi i N_f\left[\frac {l_1k_1}{K^2} -\frac{k_1}{K}\right]}
\epsilon^{(k_1, l_1)}_{(k_1+k_2,l_1+l_2-K)}\\
\qquad \text{if } k_1+k_2 \leq N-1, l_1+l_2 > K-1\\
\\
e^{2\pi i N_f\left[\frac {l_1k_1}{K^2} -\frac{l_1+k_1}{K}\right]}
\epsilon^{(k_1, l_1)}_{(k_1+k_2-K,l-1+l_2-K)}\\  
 \qquad \text{if } k_1+k_2, l_1+l_2 > K-1
\end{cases}
\end{align}

Let us analyze Eqs. (\ref{srule1}) and (\ref{trule1}) in general. First focus on Eq. (\ref{srule1}) and observe that when $K$ is even 
\begin{align}
 \epsilon^{(K/2,K/2)}_{(K/2,K/2)}
 = e^{\pi i N_f}\epsilon^{(K/2,K/2)}_{(K/2,K/2)},
\end{align}
so this forces $N_f$ be even. For even $K$ this is the only condition required while for odd $K$ there is no condition for consistency of Eq. (\ref{srule1}) as is easily verified. 

Now Eq.\ (\ref{trule1}) gives after few iterations
\begin{align}
\epsilon^{(1,1)}_{(0,0)} = e^{\frac{2\pi i N_f}{K^2}}\epsilon^{(1,1)}_{(1,1)}
=
\dots =e^{\frac{2\pi i N_f(K-1)}{K^2}}\epsilon^{(1,1)}_{(K-1,K-1)}\nonumber\\
=e^{-\frac{2\pi i N_f}{K}}\epsilon^{(1,1)}_{(0,0)},
\end{align}
which is consistent iff $N_f = 0$ mod $K$. 

For generic values of $k_1,k_2,l_1,l_2$ one obtains the following consistency condition
\begin{align}
 \epsilon^{(k_1,l_1)}_{(k_2,l_2)} = e^{\frac{2\pi i N_fl_1k_1p}{K^2}} 
 \epsilon^{(k_1,l_1)}_{(k_2,l_2)} ,
\end{align}
for some integers $p,s,t$ such that $pk_1 - Kt = 0$ and $pl_1 -Ks=0$. Therefore $T$ modular invariance is possible iff $N_f = 0$ mod $K$. It is not difficult to show that with this condition on $N_f$ and the phases, 
$S$ and $T$ modular invariance can be simultaneously achieved. Hence we conclude that modular invariance is possible iff $N_f= 0$ mod $K$.

\subsubsection{Example: {$\mathbb Z_2 \times \mathbb Z_2$}}

Equation (\ref{srule1}) gives 
$\epsilon^{(1,1)}_{(1,1)} = e^{i N_f \pi} \epsilon^{(1,1)}_{(1,1)}$ which is 
possible, 
for non-zero $\epsilon^{(1,1)}_{(1,1)}$,
iff $N_f= 0$ mod 2. 
With these conditions (\ref{srule1}) and (\ref{trule1}) give
\begin{align}
 \epsilon^{(0,0)}_{(0,1)} &= \epsilon^{(0,1)}_{(0,0)}=\epsilon^{(0,1)}_{(0,1)},
\quad
\epsilon^{(0,0)}_{(1,0)} = \epsilon^{(1,0)}_{(0,0)} = \epsilon^{(1,0)}_{(1,0)},
\nonumber \\
\epsilon^{(0,0)}_{(1,1)} &= \epsilon^{(1,1)}_{(0,0)}=\epsilon^{(1,1)}_{(1,1)},
\nonumber\\
\epsilon^{(0,1)}_{(1,0)}&
=\epsilon^{(0,1)}_{(1,1)}
=\epsilon^{(1,1)}_{(1,0)}=\pm\epsilon^{(1,1)}_{(0,1)}
= \pm\epsilon^{(1,0)}_{(0,1)}
=\pm\epsilon^{(1,0)}_{(1,1)}, 
\end{align}
with the minus signs when $N_f = 2$ mod 4 and plus signs when $N_f = 0$ mod 4.
Therefore modular invariance can be achieved iff $N_f = 0$ mod 2.
That is the edge theory is expected to be unstable when
$N_f=0$ mod 2. 
In Sec.\ 
\ref{gapping potential}, 
we will construct explicitly a potential
that gaps out the edge theory.

\subsection{$\mathbb Z_K$ symmetry}

To compare our results to that obtained
for $\mathbb Z_K$ SPT phases, 
\cite{LuVishwanath2012}
we consider the embedding $\mathbb Z_K \rightarrow \mathbb Z_K\times \mathbb Z_K$ given by $k \rightarrow (k, kq)$ and $N_f=1$. To analyze this case we simply set $l_1 = k_1q$ and $l_2 = k_2q$ and $N_f=1$ in the previous expressions above. Hence we can write
\begin{align}
 Z(\tau)  = \sum_{k_1,k_2=0}^{K-1} \epsilon^{k_1}_{k_2}Z^{k_1}_{k_2}(\tau)
\end{align}
with $Z^{k_1}_{k_2} := Z^{(k_1, k_1q)}_{(k_2, k_2q)}$ and similarly for $\epsilon$. Therefore in this case the large gauge anomaly is
\begin{align}
 Z^{k_1}_{k_2\pm K} 
= Z^{(k_1, qk_1)}_{(k_2 \pm K, k_2q \pm qK)}
= e^{\pm \frac{4\pi i q k_1}{K}} Z^{k_1}_{k_2}. 
\end{align}
The $T$ modular transformation is now given by 
\begin{equation}\label{ttransformation2}
 Z^{k_1}_{k_2}(\tau+ 1 ) = e^{-\frac{2\pi i qk_1^2}{K^2}} Z^{k_1}_{k_1+k_2}(\tau), 
\end{equation}
while the $S$ modular transformation is 
\begin{equation}\label{srule2}
 Z^{k_1}_{k_2} (-1/\tau) = e^{\frac{4\pi i qk_1k_2}{K^2}} Z^{k_2}_{-k_1} (\tau).
\end{equation}
Thus the $S$ modular invariance condition  is 
\begin{align}\label{srule2f}
 \epsilon^{k_1}_{k_2} = 
\begin{cases}
 e^{-\frac{4\pi i qk_1k_2 }{K^2}} \epsilon^{k_2}_{-k_1}
 \quad
\text{if}
\quad k_1=0,
\\
\\
e^{-4\pi i (\frac{qk_1k_2}{K^2} -\frac{qk_2}{N})} 
\epsilon^{k_2}_{-k_1+K}
 \quad \text{if}\quad  k_1 > 0,
\end{cases}
\end{align}
while the $T$ modular invariance condition is 
\begin{align}\label{trule2}
 \epsilon^{k_1}_{k_2} = 
\begin{cases}
 e^{\frac{2\pi i q{k_1}^2 }{K^2}} \epsilon^{k_1}_{k_1+k_2}
& \text{if}\quad  k_1+k_2\leq K-1,
\\
\\
e^{2\pi i\left[\frac {q{k_1}^2}{K^2} -\frac{2qk_1}{K}\right]}
 \epsilon^{k_1}_{k_1+k_2-K} 
 & \text{if} \quad k_1+k_2 > K-1. 
\end{cases}
\end{align}

From Eq.\ (\ref{srule2f}) above,
one obtains
\begin{align}
 \epsilon^0_{k_2} = \epsilon^{k_2}_{0} 
 = \epsilon^0_{-k_2 +K}
 = \epsilon^{-k_2+K}_{0}, 
\end{align}
if $k_1 = 0$ and $k_2 \neq 0$, 
whereas
\begin{align}
 \epsilon^{k_1}_{0} = \epsilon^0_{-k_1+K} = 
 \epsilon^{-k_1+K}_{0} = \epsilon^{0}_{k_1}. 
\end{align}
if $k_2 = 0$ and $k_1\neq 0$.
Similarly, for even $K$ and $k_1 = k_2 = K/2$ one gets
\begin{align}
 \epsilon^{K/2}_{K/2}= e^{\pi i q}
 \epsilon^{K/2}_{K/2}
\end{align}
which is consistent iff $q = 0$ mod $2$. 
Finally, 
for other values of $k_1$ and $k_2$ different from those considered above one gets
\begin{align}
 \epsilon^{k_1}_{k_2} 
&= e^{-4\pi i(\frac{qk_1k_2}{K^2}-\frac{qk_2}{K})}
\epsilon^{k_2}_{-k_1+K}
\nonumber\\
&= e^{-\frac{4\pi i qk_1}{K} }\epsilon^{-k_1+K}_{-k_2+K}
\nonumber\\
&=e^{-\frac{4\pi i qk_1k_2}{K^2} }\epsilon^{-k_2+K}_{k_1}
\end{align}
where we have left out a self consistent further iteration.
This shows that $S$ modular invariance is possible for even $K$ iff $q$ is even and for odd $K$ there is no condition on $q$.

Now we move on to analyze the $T$ modular invariance conditions 
which imply
\begin{align}
 \epsilon^1_0 = 
 e^{\frac{2\pi i q}{K^2}} \epsilon^{1}_{1}
 &=
 \dots 
 =e^{\frac{2\pi i q(K-1)}{K^2}} \epsilon^{1}_{K-1}
 =e^{-\frac{2\pi i q}{K}} \epsilon^{1}_{0}, 
\end{align}
which is self consistent iff $q= 0$ mod $K$. 
We now show that this condition is sufficient for $T$ modular invariance.
When $k_1 \neq 0$ after several reiterations one gets the condition
\begin{align}\label{tphase2}
 \epsilon^{k_1}_{k_2} = e^{\frac{2\pi i q k_1^2 p}{K^2}} 
 \epsilon^{k_1}_{k_2}, 
\end{align}
 where $p$ is an integer such that $k_1p - Kt=0$ for an integer $t$ whose actual value is irrelevant to us. So that with $q= 0$ mod $K$ the phase in
 Eq.\ (\ref{tphase2}) is just $1$. 

Thus putting our results together we find that modular invariance is possible iff $q=0$ mod $K$.
I.e., the phase is trivial when $q=0$ mod $K$
and non-trivial (i.e., SPT phase) otherwise.

\subsection{Gapping potential perspective}

\label{gapping potential}

In this section we show that there exist potentials 
that can fully gap our system without explicitly or spontaneously 
breaking the $\mathbb Z_2 \times \mathbb Z_2$ symmetry 
iff $N_f= 0$ mod 2. 
Thus we confirm that we do indeed have a SPT phase 
when $N_f \neq 0$ mod 2 and a trivial phase otherwise.

Let us first consider the case when $N_f =1$, 
in this case a complete set of local operators in the field theory 
is given by 
$\partial \Phi$ , 
$\partial \Theta$,
and $\cos( m \Phi + n\Theta + \alpha)$, 
where $\partial$ denotes a generic derivative and $\alpha$ is a constant. 
Here we have switched our notation,
$\phi_1 \rightarrow \Phi$ and $\phi_2 \rightarrow \Theta$,
to emphasize the dual (canonical conjugate) nature of these fields,
$[\Phi, \partial \Theta]\sim 2\pi i$. 
The most general gapping potential terms that is $\mathbb Z_2 \times \mathbb Z_2$ symmetric is given by linear combinations of the form
\begin{align}
 \label{cosine}
\cos(2m \Phi + 2n \Theta + \alpha).
\end{align}

Now we can do a semi-classical analysis to show that 
the ground state spontaneously breaks 
$\mathbb Z_2 \times \mathbb Z_2$ symmetry
once a strong enough gapping potential
of the form (\ref{cosine})  
is added. 
Without loss of generality we set $\alpha = 0$,  
and since $[\Phi, \partial\Theta] \sim 2\pi i$ etc.,
we need to consider independently potentials of the form $\cos(2m\Phi)$ and $\cos(2n \Theta)$. For $\cos(2m\Phi)$ we have that classical minima correspond to $\Phi = \frac {2j+1}{2m} \pi$ 
(for a finite number of independent $j$'s since $\Phi \in [0, 2\pi)$) and the $\mathbb Z_2$ transformation $\Phi \rightarrow \Phi + \pi$ amounts to $j \rightarrow j + m$. As one can easily see when $m = 1$, for example, these classical minima transform under $\mathbb Z_2$  and a choice of any one of them would spontaneously break $\mathbb Z_2$ symmetry. The analysis for potentials of the form $\cos(2n \Theta)$ is similar and we reach the conclusion that when $N_f =1$ it is not possible to gap our system without breaking $\mathbb Z_2 \times \mathbb Z_2$ symmetry. 

When $N_f = 2$ we have fields 
$\Phi_1, \Theta_1, \Phi_2, \Theta_2$ and the most general $(\mathbb Z_2 \times \mathbb Z_2)$-invariant possible gapping potential is given by
$\cos
[m_1(\Phi_1 - \Phi_2) + n_1(\Theta_1-\Theta_2) + 2 l_1 \Phi_1 + 2 l_2 \Theta_2 + \alpha]
$
where $m_1, n_1, l_1, l_2\in \mathbb{Z}$. 
If we focus on two mutually commuting
and $(\mathbb{Z}_2\times\mathbb{Z}_2)$-symmetric
bosonic fields $\Phi_1-\Phi_2$
we can consider a gapping potential
$
\cos(m_1(\Phi_1-\Phi_2)+n_1(\Theta_1+\Theta_2))
$.
If we consider $-\cos(\Phi_1-\Phi_2) -\cos (\Theta_1+\Theta_2)$ we see that the classical minima corresponds to
\begin{align}
\Phi_1-\Phi_2 = 0,
\quad 
\Theta_1 + \Theta_2 = 2\pi ,
\end{align}
since this is invariant under $\mathbb Z_2 \times \mathbb Z_2$ and $[\Phi_1-\Phi_2, \Theta_1 + \Theta_2] = 0$ we conclude that we can fully gap the system without spontaneously  breaking the symmetry.

Most generally when  $N_f = 2m$ we have fields $\Phi_i, \Theta_i$ 
($i = 1, \dots, 2m$) and the most general $(\mathbb Z_2 \times \mathbb Z_2)$ invariant possible gapping potential is given by a sum over potentials of the form 
\begin{align}
\cos\Big\{\sum_{i=1}^{2m-1} m_i(\Phi_i - \Phi_{2m})  +\sum _{i =1}^{2m-1}n_i(\Theta_i-\Theta_{2m}) 
\nonumber\\
+ 2 l_1 \Phi_{2m} + 2 l_2 \Theta_{2m} + \alpha \Big\}.
\end{align}
 
If we focus on two mutually commuting
and $(\mathbb{Z}_2\times\mathbb{Z}_2)$-symmetric
bosonic fields then a potential of the form
\begin{align}
-\sum_{i =1}^m
\left[\cos(\Phi_{2i-1} - \Phi_{2i})+  \cos(\Theta_{2i-1} +\Theta_{2i})
\right] 
\end{align}
would be minimized classically by 
\begin{align} \label{gap}
\Phi_{2i-1} - \Phi_{2i}= 0,
\quad
\Theta_{2i-1} +\Theta_{2i}= 2\pi, 
\end{align}
which is allowed since all the commutators between the different fields on the left-hand side (LHS) of (\ref{gap}) are zero. On the other hand when 
$M= 2m+1$, by choosing a maximal set of $2m$ commuting $(\mathbb Z_2 \times \mathbb Z_2)$-invariant fields (basically differences $\Phi_i - \Phi_ j$ etc) to localize, one is always left with one field that cannot be localized and hence the system cannot be fully gapped. This concludes our analysis of bosonic SPT phases with  $\mathbb Z_2 \times \mathbb Z_2$  symmetry.

For general $K$ 
the above analysis is easily generalized. 
For example, for $K=3$ with $N_f = 3$ one can gap out
the following independent, mutually commuting, and $(\mathbb Z_K \times \mathbb Z_K)$-invariant combinations
\begin{align}
\Phi_1-\Phi_2, \hspace{3mm}\Phi_2-\Phi_3,\hspace{3mm} \Theta_1+\Theta_2+\Theta_3,
\end{align}
while for $K =4$ with $N_f =4$ one can gap out
\begin{align}
\Phi_1-\Phi_2, \hspace{3mm}\Theta_1+\Theta_2-\Theta_3-\Theta_4,\nonumber\\
\Phi_3-\Phi_4,\hspace{3mm}\Theta_1+\Theta_2+\Theta_3+\Theta_4.
\end{align}
For generic $K$, one could consider, for example,  
a gapping potential  
$\sum_{i=1}^{K-1}\cos[m_i(\Phi_{i}-\Phi_{i-1})]
+\cos[n\sum_i^{K}\Theta_{i}]
$
that gaps $K=N_f$ mutually independent and $(\mathbb{Z}_K \times \mathbb{Z}_K)$
symmetric combinations of bosonic fields. 

\section{
Fermionic symmetry protected topological phases}

In this section,
we are interested in fermionic SPT phases which are relevant 
to the physics of topological superconductors. 
We consider $\mathbb Z_K \times \mathbb Z_K$ symmetry which is a generalization of the $\mathbb Z_2 \times \mathbb Z_2$ symmetry discussed in Refs. \onlinecite{RyuZhang2012, LiangQi}. 
The relevant edge theory in the absence of interactions 
with several flavors is described by the free Dirac 
action
\begin{equation}\label{fermionedge}
S= \frac{i}{2\pi} \int dx dt\,
\sum^{N_f}_{a=1}
\psi^\dagger_{La}
(\partial_t-\partial_x) 
\psi_{La}
+  
\psi^\dagger_{Ra}(\partial_t+\partial_x) \psi_{Ra}. 
\end{equation}
where $\psi^{\dag}_L, \psi^{\ }_L, \psi^{\dag}_R, \psi^{\ }_R$
are creation/annihilation operators of the fermionic 
nonchiral edge modes
that are supported by some topological bulk system. This action has a $U(1) \times U(1)$ symmetry,
which contains $\mathbb{Z}_K\times \mathbb{Z}_K$ as its 
subgroup. 
As in the bosonic case we gauge 
this $\mathbb Z_K \times \mathbb Z_K$ subgroup
to understand the corresponding SPT phases.

 \subsection{Partition function and modular transformation}
The Hamiltonian $H$ and momentum $P$,
when $N_f=1$, are
\begin{align}
H &= \frac{i}{2\pi} \int dx 
\left[  \psi^\dagger_L\partial_x\psi_L -\psi^\dagger_R\partial_x\psi_R
\right], 
\nonumber\\
P &= \frac{i}{2\pi} \int dx
\left[ 
 \psi^\dagger_L\partial_x\psi_L +\psi^\dagger_R\partial_x\psi_R
\right].
\end{align}
There is an obvious global $U(1) \times U(1)$ symmetry under the transformations
\begin{align}
\psi_L \rightarrow e^{2\pi i\alpha}\psi_L, 
\quad  
\psi_R\rightarrow e^{2\pi i\tilde \alpha}\psi_R
\end{align}
with the left movers neutral under the second $U(1)$ factor and the right movers neutral under the first $U(1)$ factor. This is generated by the left and right ``fermion number'' charges
\begin{align}
 Q_L = \frac{1}{2\pi} \int dx\, \psi^\dagger_L\psi^{\ }_L,
\quad 
Q_R = \frac{1}{2\pi} \int dx\,  \psi^\dagger_R\psi^{\ }_R.
\end{align}
The total fermion number is 
$Q:=Q_L+Q_R$.
We work on a complex torus of modulus $\tau$ and consider sectors with 
the following periodicity conditions:
\begin{align}\label{periodicity}
\psi_L(x+2\pi,t) &= e^{-2\pi i\alpha}\psi_L(x,t),
\nonumber \\
\psi_L(x+2\pi\tau_1, t+2\pi\tau_2) &= e^{-2\pi i\beta}\psi_L(x,t),
\nonumber \\
\psi_R(x+2\pi,t) &= e^{2\pi i\tilde\alpha}\psi_R(x,t),
\nonumber \\
\psi_R(x+2\pi\tau_1, t+2\pi\tau_2) &= e^{2\pi i\tilde\beta}\psi_R(x,t).
\end{align}
(In the operator formalism, 
we can account for the spatial periodicity by 
appropriate mode expansions 
and 
for the time periodicity by insertion of an appropriate operator.) 
The mode expansions are
\begin{align}
 \label{mode}
\psi_L(x,t) = \sum_{r \in \mathbb Z+ \alpha} a_r e^{-ir(x+t)},
\nonumber\\
\psi_R(x,t) = \sum_{r \in \mathbb Z+ \tilde\alpha} b_r e^{-ir(t-x)}, 
\end{align}
with anticommutators
\begin{equation}
\{a_r, a^\dagger_s\} = \{b_r, b^\dagger_s\} =\delta_{r,s}.
\end{equation}
%
We introduce the normal ordering with respect to
the fermionic ground state as
\begin{align}
    :a_r^\dagger a_r:= 
\begin{cases}
    a_r^\dagger a_r &\text{if } r\geq 0,
    \\
    -a_r a_r^\dagger &\text{if} \hspace{2mm}  r < 0, 
\end{cases}
\end{align}
and similarly for the $b$ oscillators. 
So the left-moving contribution to $H$,
$H_L=\sum_{r\in \mathbb{Z}+\alpha}r a^{\dag}_r a^{\ }_r$, 
is now given by
\begin{align}
H_L 
= 
\sum_{r \in \mathbb Z+ \alpha} r:a_r^\dagger a_r:  
 -\frac{1}{24} + \frac{1}{2}(\alpha-[\alpha] -\frac{1}{2})^2, 
\end{align}
where we have adopted the regularization prescription 
$\sum_{r < 0}r\nonumber = E_0(\alpha) = -1/24 + ({1}/{2})(\alpha-[\alpha] -{1}/{2})^2
$
and $[\alpha]$ is the greatest integer less 
than or equal to $\alpha$.
We get a similar contribution from the 
right movers with $\alpha$ replaced by $\tilde \alpha$,
$H_R= \sum_{s\in \mathbb{Z}+\tilde{\alpha}} s b^{\dag}_s b^{\ }_s$.
The left moving contribution to the fermion number,
$Q_L=\sum_{r\in \mathbb{Z}+\alpha} a^{\dag}_r a^{\ }_r$, 
is
\begin{align}
Q_L 
&= \sum_r :a_r^\dagger a_r: + \alpha -[\alpha] -\frac{1}{2}
\end{align}
with the regularization prescription
$\sum_{r < 0} 1 = \alpha-[\alpha]-1/2$.
We get a similar contribution from the right movers with $\alpha$ replaced by $\tilde \alpha$,
$Q_R=\sum_{s\in \mathbb{Z}+\tilde{\alpha}}b^{\dag}_s b^{\ }_s$.

We would like to evaluate the partition function in Euclidean signature for fermions with boundary conditions (\ref{periodicity}).
This is given by ($q:= e^{2\pi i \tau}$) 
\begin{align}
&Z^{\alpha \tilde\alpha}_{ \beta \tilde \beta}(\tau) 
=
\mathrm{Tr}_{\alpha}
\left[ e^{2\pi i Q_L (-\beta + \frac{1}{2} )}q^{H_L}
\right]
\nonumber \\
&\qquad 
\times 
\mathrm{Tr}_{\tilde\alpha}
\left[e^{-2\pi i Q_R (-\tilde\beta + \frac{1}{2} )} 
 \bar{q}^{ H_R}
\right]
= Z^\alpha_\beta(\tau) \bar Z^{\tilde \alpha}_{\tilde\beta}(\tau),
\end{align}
where the trace is taken over the fermionic Fock space 
generated by the fermonic oscillator modes  
in (\ref{mode}) 
acting on the Dirac sea. 
So the partition function has a left-right factorized form. We focus on the left movers since the analysis is similar for the right movers.
\begin{align}\nonumber
&\quad Z^\alpha_\beta (\tau) 
\nonumber \\
&= 
\mathrm{Tr}_{\alpha} \left[
 e^{2\pi i(\sum_r :a_r^\dagger a_r: + \alpha-[\alpha]-\frac{1}{2})(-\beta +\frac{1}{2})} 
 q^{\sum_r r:a_r^\dagger a_r: + E_0(\alpha)}\right]
\nonumber\\
&= e^{2\pi i(\alpha-[\alpha]-\frac{1}{2})(-\beta + \frac{1}{2})}
\prod_{r \geq 0}
\left[1+ e^{2\pi i (-\beta +\frac{1}{2})}q^r\right]
\nonumber\\ 
&\quad
\times q^{E_0(\alpha)}\prod_{r< 0} 
\left[1+ e^{-2\pi i (-\beta +\frac{1}{2})}q^{-r}\right]. 
\end{align}
Note that $Z^{\alpha+1}_\beta(\tau) =Z^{\alpha}_\beta(\tau)$ and $Z^{\alpha}_{\beta+1}(\tau) = e^{-2\pi i(\alpha-[\alpha]-\frac{1}{2})} Z^{\alpha}_\beta(\tau)$. 
We can write the partition function as 
\begin{align}
 \label{fermionpartition}
\quad Z^\alpha_\beta(\tau)
&=
e^{\pi i \tau ([\alpha]^2-2(\alpha-\frac{1}{2})[\alpha])-2\pi i[\alpha](-\beta +\frac{1}{2})}
\nonumber \\
&\quad \times 
 \frac{\theta \begin{bmatrix} \alpha-\frac{1}{2} \\ -\beta +\frac{1}{2}
 \end{bmatrix}(0,\tau)}{\eta(\tau)},  
\end{align}
where the theta function with characteristics is given by
\begin{align}
\theta \begin{bmatrix} a\\ b\end{bmatrix}(\nu,\tau) = \sum_{n \in \mathbb Z} e^{\pi i (n+a)^2 \tau + 2\pi i(n+a)(\nu +b)}.  
\end{align}
If we choose $\alpha \in [0,1)$ then $[\alpha] = 0$ 
and the phase factor in Eq.\ (\ref{fermionpartition}) cancels.


By noting 
\begin{align}
\theta \begin{bmatrix} a\\ b\end{bmatrix}(\nu,\tau+1)= e^{-\pi i(a^2+a)} \theta \begin{bmatrix} a\\ b+a+\frac{1}{2}\end{bmatrix}(\nu,\tau),
\end{align}
and using the Poisson resummation formula 
one finds  
\begin{align}
\theta \begin{bmatrix} a\\ b\end{bmatrix}(\nu,-\frac{1}{\tau})= (-i\tau)^{\frac{1}{2}}e^{2\pi i a(\nu +b)} \theta \begin{bmatrix} -\nu-b\\ a\end{bmatrix}(0,\tau).
\end{align}
%
From these one deduces the following modular transformations of the right moving contribution to the partition function:
\begin{align}
Z^{\alpha}_{\beta}(\tau+1) &= e^{-\pi i(\alpha^2-\frac{1}{6})}Z^\alpha_{\beta - \alpha}(\tau),
\nonumber \\
Z^{\alpha}_{\beta}(-1/\tau) &= e^{2\pi i (\alpha-\frac{1}{2})(-\beta+\frac{1}{2})}Z^\beta_{1-\alpha}(\tau), 
\end{align}
where we have assumed $\alpha, \beta \in [0,1)$ as would be the relevant case in what follows.\\

\subsection{
$\mathbb Z_K \times \mathbb Z_K$ symmetry
}

We are interested in gauging $\mathbb Z_K \times \mathbb Z_K$ symmetry of $N_f$ Dirac fermions ($2N_f$ Majorana fermions) which is generated by
\begin{align}
\psi_L^a \rightarrow e^{-\frac{2\pi i }{K}}\psi_L^a
\quad \mbox{and}\quad 
\psi_R^a \rightarrow e^{\frac{2\pi i }{K}}\psi_R^a
\end{align}
where $a = 1, \dots, N_f$.
Thus we shall be analyzing the modular properties of the most general partition function
\begin{align}
Z (\tau)= \sum_{k, l,m,n} 
\epsilon_{(k,l,m,n)}
\left[
 Z^{k'}_{l'}(\tau){\bar Z}^{m'}_{n'}(\tau)
\right]^{N_f}, 
\end{align}
with $k,l,m,n = 0,1,\dots, K-1$, and $k' := {k}/{K}$ etc.
This corresponds to a sum over the most general combination of left and right twistings of the spinors by the $\mathbb Z_K \times \mathbb Z_K$ action.
With the $\epsilon_{(k,l,m,n)}$ phases, 
topologically distinct sectors in the path integral sum 
are weighted differently. 
The modular transformations are given by
\begin{align}
 \label{ttransformation2}
Z(\tau+1)  &= 
\sum_{k,l,m,n} \epsilon_{(k,l,m,n)} e^{-\pi i N_f(\frac{k^2-m^2}{K^2})}
\\
&
\quad \times 
\left[
 Z^{k'}_{l'-k'}(\tau) \bar Z^{m'}_{n'-m'}(\tau)
\right]^{N_f}, 
\nonumber \\
\label{stransformation2}
Z(-1/\tau) &=
\sum_{k,l,m,n}  \epsilon_{(k,l,m,n)}  e^{2\pi i N_f(\frac{mn-kl}{K^2}+\frac{k-m+n-l}{2K})}\nonumber\\
&\quad \times 
\left[
 Z^{l'}_{1-k'}(\tau) \bar Z^{n'}_{1-m'}(\tau)
\right]^{N_f}. 
\end{align}

As in the bosonic case, we will ask when the modular invariance
can be achieved for different values of $K$ and $N_f$. 
Instead of being exhaustive, we make an ansatz 
that the overall partition function is holomorphically factorized so that
\begin{equation}
Z(\tau) = Z_L(\tau) \bar Z_R(\tau) = |Z_L(\tau)|^2. 
\end{equation}
In this case modular invariance is achieved as long as the left(right)-moving contribution transforms covariantly with a phase under a generic modular 
transformation $U$,
$Z_L(U\tau) = e^{i\theta} Z_L(\tau)$. 
So we can focus on the left-moving sector, 
\begin{equation}
Z_L(\tau) = \sum_{k,l = 0}^{K-1}
\epsilon^k_l  Z_{l'}^{k'}(\tau). 
\end{equation}

The condition for $S$ modular invariance is then deduced from Eq.\ (\ref{stransformation2}) to be
(the modular invariance conditions presented below are all up to an overall phase)
\begin{align}\label{srule2}
 \epsilon^k_l = 
 \begin{cases}
  e^{2\pi i N_f(\frac{kl}{K^2}+\frac{l-k}{2K})} \epsilon^{l}_{K-k}
& \text{if } k>0\\
\\
e^{2\pi i N_f(\frac{3l}{2K})}\epsilon^{l}_{-k} 
& \text{if } k=0\\
\end{cases}
\end{align}
while from (\ref{ttransformation2}) the conditions for $T$ modular invariance is found to be
\begin{align}\label{trule2}
 \epsilon^{k}_{l} = 
\begin{cases}
 e^{\frac{\pi i N_fk^2}{K^2}}\epsilon^{k}_{l-k}
& \text{if } 0\leq l-k \leq K-1\\
\\
e^{2\pi i N_f\left[\frac{k^2}{2K^2} +\frac{1}{2}-\frac{k}{K}\right]}
\epsilon^{k}_{l-k+K} 
& \text{if } l-k <0
\end{cases}
\end{align}

Let us analyze Eq.\ (\ref{srule2}) in general. When $K$ is even one gets
\begin{align}
 \epsilon^{k/2}_{k/2}  
 = e^{\pi i N_f/2}\epsilon^{k/2}_{k/2} 
\end{align}
so this requires $N_f =0$ mod 4. One can check that this condition is enough for self consistency of 
Eq.\ (\ref{srule2}) while for odd $K$, no condition on $N_f$ is required. 
Equation (\ref{trule2}) gives after $K$ iterations 
\begin{align}
 \epsilon^{1}_{0} = e^{\pi i N_f\frac{K-1}{K}}
 \epsilon^{1}_{0}. 
\end{align}
When $K$ is even this requires $N_f= 0$ mod $2K$ while 
for odd $K$ this requires $N_f = 0$ mod $K$.
In general several iterations of Eq.\ (\ref{trule2}) gives
\begin{align} \label{const}
 \epsilon^{k}_{l}= 
 e^{\pi iN_f(\frac{k^2p}{K^2}+t-\frac{2kt}{K})}
 \epsilon^{k}_{l}, 
\end{align}
for some integers $p$ and $t$ such that 
\begin{align}\label{integers}
Kt -kp = 0. 
\end{align}
For even $K$ one sees that $N_f= 0$ mod $2K$ 
is sufficient while for odd $K$ with $N_f = 0$ mod $K$ the phase in (\ref{const}) is $e^{\pi i N_f t\frac{k+K}{K}}$. 
So one needs to consider only the case when $k$ is even. 
In this case (\ref{integers}) implies that 
$t$ must be even. Hence the phase is one and we conclude that modular invariance is possible for even $K$ iff $N_f = 0$ mod $2K$ while for odd 
$K$ iff $N_f =0$ mod $K$.

\subsection{$\mathbb Z_2 \times \mathbb Z_2$ Example}

For the case $\mathbb Z_2 \times \mathbb Z_2$, 
let us check that we recover the results in 
Ref.\ \onlinecite{RyuZhang2012} from our analysis. 
In this case Eq.\ (\ref{srule2}) gives
\begin{align} \label{sz2rule}
 \epsilon^{0}_{1} = 
 e^{\frac{3\pi i N_f}{2}} \epsilon^{1}_{0},
\quad
\epsilon^{1}_{0} =  e^{-\frac{\pi i N_f}{2}} 
\epsilon^{0}_{1},
\quad
\epsilon^{1}_{1} = e^{\frac{\pi i N_f}{2}} 
\epsilon^{1}_{1}, 
\end{align}
while Eq.\ (\ref{trule2}) gives
\begin{align}\label{tz2rule}
 \epsilon^{1}_{0} = e^{-\frac{3\pi i N_f}{4}} 
 \epsilon^{1}_{1},
\quad
\epsilon^{1}_{1} =e^{\frac{\pi i N_f}{4}} 
\epsilon^{1}_{0}.
\end{align}
Clearly Eqs.\ (\ref{sz2rule}) and (\ref{tz2rule}) are consistent 
iff $N_f = 0$ mod 4 which recovers the result in 
Ref.\ \onlinecite{RyuZhang2012} with $N_f = 2N$ being the number of flavors of Majorana modes. This result is, of course, familiar from type II string theory \ \cite{Polchinski98} where the 
$\mathbb{Z}_2$ symmetry is generated by the fermion number current on the world sheet and the Majorana modes corresponds to directions in spacetime with two extra dimensions cancelling the ghosts modes that result from gauge fixing. Thus the GSO left right assymmetric $\mathbb Z_2$ projection gives rise to consistent modular invariant superstring theories in $8+2$ spacetime dimensions.

\subsection{Gapping potential perspective}

To analyze the stability to interactions it is convenient to bosonize the fermionic fields as follows:
\begin{align}
\psi_L &= Ae^{i \phi_L},\quad 
\psi_R = Ae^{i\phi_R}, 
\end{align}
where an implicit normal ordering has been omitted on the right hand side. We would also make use of 
\begin{align}
2\phi_L = \phi+ \theta, 
\quad 
2\phi_R = \phi -\theta, 
\end{align}
so that the $\mathbb Z_K \times \mathbb Z_K$ symmetry is generated by
\begin{align}
\phi_L \rightarrow \phi_L - \frac{2\pi}{K},
\quad \phi_R \rightarrow \phi_R +\frac{2\pi}{K}. 
\end{align}

Let us first consider the simplest case of $\mathbb Z_2 \times \mathbb Z_2$. With four  flavors one can write down the interaction term
\begin{align}
V_2 = \left(\psi_{R,1}^\dagger\psi_{R,2}^\dagger\psi_{L,3}^\dagger\psi_{L,4}^\dagger+
\psi_{R,1}^\dagger\psi_{R,3}^\dagger\psi_{L,2}^\dagger\psi_{L,4}^\dagger\nonumber\right.\\
\left.+\psi_{R,1}^\dagger\psi_{R,4}^\dagger\psi_{L,2}^\dagger\psi_{L,3}^\dagger+ L \leftrightarrow R \right)+ c.c. 
\end{align}
In the bosonized form this is just
\begin{align}
V_2 &= 4A^4 \cos\tilde\phi_1\left(\cos\tilde \theta_2 +\cos\tilde \theta_3+\cos\tilde \theta_4\right)
\end{align}
where 
\begin{align}\label{redef}
2\tilde\phi_1 &= \phi_1 +\phi_2 +\phi_3 +\phi_4,
\nonumber\\
2\tilde\theta_2 &= \theta_1+\theta_2-\theta_3-\theta_4,
\nonumber\\
2\tilde\theta_3 &= \theta_1+\theta_3-\theta_2-\theta_4,
\nonumber\\
2\tilde\theta_4 &= \theta_1+\theta_4-\theta_2-\theta_3, 
\end{align}
using $[\phi_i, \theta_j] \sim i\delta_{ij}$ 
one sees that all the fields in Eq.\ (\ref{redef})
mutually commute and can be simultaneously localized. Also one sees that there are two solutions for the classical minima given by
\begin{align}
\tilde\phi_1= \pi m_1,
\quad 
\tilde\theta_2= \pi n_2,
\quad 
\tilde\theta_3= \pi n_3, 
\quad 
\tilde\theta_4= \pi n_4,
\end{align}
with either $m_1$ odd and $n_2,n_3,n_4$ even or $m_1$ even and $n_2,n_3,n_4$ odd. By using the periodicity $\phi_i = \phi_i + 2\pi$ and similarly for $\theta_i$ one sees that these two solutions are equivalent and the invariance of $\tilde\phi_1,\tilde\theta_2,\tilde\theta_3,\tilde\theta_4$ implies that the system can be fully gapped without explicitly or spontaneously breaking $\mathbb Z_2 \times \mathbb Z_2$ symmetry.

Let us now consider the case of 
$\mathbb Z_4 \times \mathbb Z_4$ symmetry
as this would enable us to understand the generalization to $\mathbb Z_{2K} \times \mathbb Z_{2K}$. In this case with $N_f = 8$ one can write down 
the following potential 
\begin{align}
V_4 &= \psi^{\dagger}_{R,1}\psi^{\dagger}_{R,2}\psi^{\dagger}_{R,3}\psi^{\dagger}_{R,4}
\psi^{\dagger}_{L,5}\psi^{\dagger}_{L,6}\psi^{\dagger}_{L,7}\psi^{\dagger}_{L,8}
\nonumber\\
& \quad
+\psi^{\dagger}_{R,1}\psi^{\dagger}_{R,2}\psi^{\dagger}_{R,6}\psi^{\dagger}_{R,8}
\psi^{\dagger}_{L,3}\psi^{\dagger}_{L,4}\psi^{\dagger}_{L,5}\psi^{\dagger}_{L,7}
\nonumber\\
&\quad
+\psi^{\dagger}_{R,1}\psi^{\dagger}_{R,2}\psi^{\dagger}_{R,5}\psi^{\dagger}_{R,6}
\psi^{\dagger}_{L,3}\psi^{\dagger}_{L,4}\psi^{\dagger}_{L,7}\psi^{\dagger}_{L,8}
\nonumber\\
&\quad
+\psi^{\dagger}_{R,1}\psi^{\dagger}_{R,3}\psi^{\dagger}_{R,5}\psi^{\dagger}_{R,7}
\psi^{\dagger}_{L,2}\psi^{\dagger}_{L,4}\psi^{\dagger}_{L,6}\psi^{\dagger}_{L,8}
\nonumber\\
&\quad
+\psi^{\dagger}_{R,1}\psi^{\dagger}_{R,3}\psi^{\dagger}_{R,6}\psi^{\dagger}_{R,8}
\psi^{\dagger}_{L,2}\psi^{\dagger}_{L,4}\psi^{\dagger}_{L,5}\psi^{\dagger}_{L,7}
\nonumber\\
&\quad
+\psi^{\dagger}_{R,1}\psi^{\dagger}_{R,4}\psi^{\dagger}_{R,5}\psi^{\dagger}_{R,8}
\psi^{\dagger}_{L,2}\psi^{\dagger}_{L,3}\psi^{\dagger}_{L,6}\psi^{\dagger}_{L,7}
\nonumber\\
&\quad
+\psi^{\dagger}_{R,1}\psi^{\dagger}_{R,4}\psi^{\dagger}_{R,6}\psi^{\dagger}_{R,7}
\psi^{\dagger}_{L,2}\psi^{\dagger}_{L,3}\psi^{\dagger}_{L,5}\psi^{\dagger}_{L,8}. 
\end{align}
In terms of bosonized fields this is 
\begin{align}
V_4 = 4A^8 \cos\tilde \phi_1 \left(\sum_{i =2}^8\cos\tilde \theta_i\right)
\end{align}
where
\begin{align}
2\tilde \phi_1 &= \sum_{i=1}^8 \phi_i,
\nonumber \\
2\tilde\theta_2&=\theta_1+\theta_2+\theta_3+\theta_4-\theta_5-\theta_6-\theta_7
-\theta_8,
\nonumber\\
2\tilde\theta_3&=\theta_1+\theta_2+\theta_6+\theta_8-\theta_3-\theta_4-\theta_5
-\theta_7,
\nonumber\\
2\tilde\theta_4&=\theta_1+\theta_2+\theta_5+\theta_6-\theta_3-\theta_4-\theta_7-
\theta_8,
\nonumber\\
2\tilde\theta_5&=\theta_1+\theta_3+\theta_5+\theta_7-\theta_2-\theta_4-\theta_6-
\theta_8,
\nonumber\\
2\tilde\theta_6&=\theta_1+\theta_3+\theta_6+\theta_8-\theta_2-\theta_4-\theta_5-\theta_7,
\nonumber\\
2\tilde\theta_7&=\theta_1+\theta_4+\theta_5+\theta_8-\theta_2-\theta_3-\theta_6-\theta_7,
\nonumber\\
2\tilde\theta_8&=\theta_1+\theta_4+\theta_6+\theta_7-\theta_2-\theta_3-\theta_5-\theta_8,  
\end{align}
since the tilded fields are mutually commuting they can be simultaneously localized with $\tilde\phi_1 = \pi m_1$ and $\tilde \theta_i = \pi n_i$ with $m_1$ odd and $n_1$ even, or vice versa. Since the fields with tildes are $\mathbb Z_4 \times \mathbb Z_4$ invariant we conclude that the system can be fully gapped without breaking symmetry. It is clear that this structure can be generalized to  $\mathbb Z_{2K} \times \mathbb Z_{2K}$  with $4K$ flavors, one finds the same conclusion that the system can be gapped without breaking symmetry.

\section{Discussion}
\label{Discussion}

In conclusion 
we have proposed and developed
a theoretical framework that
allows us to determine if a given (edge) CFT 
can be gapped out or not without breaking a given set of symmetries. 
It is based on the modular invariance/non-invariance of the CFT
with symmetry projection; 
it makes use of a way any 2D CFT couples to
the background geometry (complex structure of the torus) 
and hence can be applied to a wide range of systems.

There are a number of merits to our approach;
It does not rely on
the presence/absence of a conserved U(1) charge such as particle number.
Unlike topological invariants built from single-particle electron wave functions,
our method does not rely on single-particle physics and hence is 
applicable to strongly interacting systems. 
It is simpler and more convenient than
actually looking for all possible perturbations  
that can potentially gap out the edge theory
on a case -by- case basis.
For 2D SPT phases that  
have non-abelian quasiparticles there is no $K$-matrix formulation but our approach 
can be extended to such situations. 
For example, one could consider orbifolds of Wess-Zumino-Witten(WZW) models with discrete torsion.

We have demonstrated that our scheme indeed works 
for bosonic and fermionic SPT phases with
$\mathbb Z_K \times \mathbb Z_K$ or $\mathbb{Z}_K$ symmetry. 
In particular, we have checked explicitly that
for the cases when the modular invariance is achieved,
one can find an interaction potential that can gap out the
edge theory without breaking symmetry. 

The validity of our approach based on the modular invariance
is further supported by a complementary point of view proposed in  
Refs. \onlinecite{Levin 2013, LevinGu2012}.  
In various cases,
our method based on the modular invariance
and
the arguments in 
Refs.\ \onlinecite{Levin 2013, LevinGu2012} 
that makes use of the fractional statistics
in the bulk also lead to the same conditions
for the ``gappability'' of the edge theory.

One immediate generalization of our work
is to apply our method to symmetry-enriched topological phases,
i.e., topologically ordered phases that have a set of symmetries. 
For example, our calculations for Bosonic SPT phases 
can be directly generalized to the case with 
$|\det \mathrm{K}|>1$, which has ground state degeneracy. 
We have checked for a few simple cases with 
$|\det \mathrm{K}|>1$ that
when the modular invariance is achieved we can construct 
an interaction potential to gap out the edge theory.
\cite{Xiao2013}
Other interesting future work would be to consider SET phases with non-abelian 
symmetry and(or) non-Abelian statistics.



We close with a couple of comments. As discussed in the introduction, modular invariance is a global anomaly in CFT.
On the other hand, it is interesting to note that
in CFT a local anomaly associated with rescaling invariance occurs proportionately 
to the total central charge $c$.
It is also instructive to note that
in string theory conformal invariance is a constraint and $c$ is cancelled 
by working in a critical dimension. 
In condensed matter and statistical physics
applications 
conformal invariance is a real symmetry (i.e., not a constraint)
and the appearance of a local anomaly is a quantum effect which does not spoil the consistency of the theory (since there is no associated dynamical gauge degree of freedom).

-- We have focussed entirely on modular (non-)invariance on the torus. 
One may wonder if there are other constraints that come about at higher genus 
due to modular invariance and unitarity. 
Examples on torodial compactifications of string theory are explored in Ref. \onlinecite{Vafa}, 
where it is shown that modular invariance and unitarity at genus 2 enforces more 
constraints on the phases with the various possible ways of achieving modular invariance 
corresponding to elements in the second group cohomology $H^2(G, U(1))$ for a finite Abelian group $G$. 
Perhaps one can make a connection between modular non-invariance and group cohomology as well, 
in particular the third cohomology which is relevant for the classification of 2D SPT phases. 
We will not pursue this issue further here.

\acknowledgements
We would like to thank Michael Levin for useful discussion. 
SR thanks participants in Topological Phases of Matter Program  
at the Simons Center for Geometry and Physics  
for discussion. 
This work was supported in part by the National Science Foundation under grant No.\ DMR 1064319 at the University of Illinois (X.C.)


\begin{thebibliography}{99}

\bibitem{Wenbook}
X.-G.\ Wen,
\textit{Quantum Field Theory of Many-Body Systems},
(Oxford University Press, Oxford, 2004).

\bibitem{reviewQHE}
{\it The Quantum Hall Effect},
edited by R. E. Prange and S. M. Girvin (Springer, New York, 1987).

\bibitem{NayakReview}
Chetan Nayak, Steven H. Simon, Ady Stern, Michael Freedman, Sankar Das Sarma,
Rev.\ Mod.\ Phys.\ \textbf{80}, 1083 (2008).

\bibitem{reviewTIa}
M. Z. Hasan, and C. L. Kane,
Rev.\ Mod.\ Phys.\ \textbf{82}, 3045 (2010).

\bibitem{reviewTIb}
X.-L. Qi, and S.-C. Zhang,
Rev.\ Mod.\ Phys.\ \textbf{83}, 1057 (2011).


\bibitem{KaneMele}
C.\ L.\ Kane and E.\ J.\ Mele,
Phys.\ Rev.\ Lett. \textbf{95}, 146802 (2005);
Phys.\ Rev.\ Lett. \textbf{95}, 226801 (2005).

\bibitem{Bernevig05}
B.\ A.\ Bernevig and S.-C. Zhang,
Phys.\ Rev.\ Lett. \textbf{96}, 106802 (2006).

\bibitem{Bernevig06}
B.\ A.\ Bernevig, T. Hughes and S.-C. Zhang,
Science \textbf{314}, 1757 (2006).

\bibitem{Moore06}
J.\ E.\ Moore and L.\ Balents,
Phys.\ Rev.\ B \textbf{75}, 121306(R) (2007).

\bibitem{Roy3d}
R.\ Roy,
Phys.\ Rev.\ B \textbf{79}, 195322 (2009).

\bibitem{Fu06_3Da}
L.\ Fu, C.\ L.\ Kane, and E.\ J.\ Mele,
Phys.\ Rev.\ Lett. \textbf{98}, 106803 (2007).

\bibitem{Fu06_3Db}
L.\ Fu and C.\ L.\ Kane,
Phys.\ Rev.\ B \textbf{76}, 045302 (2007).

\bibitem{qilong}
X.-L. Qi, T. L. Hughes and S.-C. Zhang,
Phys.\ Rev.\ B\
\textbf{78},
195424
(2008).

\bibitem{Schnyder2008}
A. P. Schnyder, S. Ryu, A. Furusaki, and A. W. W. Ludwig,
Phys.\ Rev.\ B \textbf{78}, 195125 (2008).

\bibitem{SRFLnewJphys}
S. Ryu, A. Schnyder, A. Furusaki and A. W. W. Ludwig,
New J. Phys.
\textbf{12},
065010
(2010).

\bibitem{Kitaev2009}
A.\ Yu Kitaev, in
\textit{Advances in Theoretical Physics: Landau Memorial Conference, Chernogolovka, Russia}, 2008, edited by V. Lebedev and M. Feigel'man, AIP Conf. Proc. No. 1134 (AIP, Melville, NY, 2009), p. 22.

\bibitem{Pollmann et al}
F. Pollmann, A. M. Turner, E. Berg,  and M. Oshikawa, 
Phys.\ Rev.\ B \textbf{81}, 064439 (2010), \texttt{arXiv:0910.1811}. 

\bibitem{ChenWenGu2011}
X. Chen, Z.-C. Gu, and X.-G. Wen, Phys.\ Rev.\ B \textbf{83}, 035107 (2011)

\bibitem{Schuch et al}
N. Schuch, D. Perez-Garcia, and I. Cirac, Phys.\ Rev.\ B
\textbf{84}, 165139, (2011), \texttt{arXiv:1010.3732}

\bibitem{ChenWenGu2011b}
X. Chen, Z.-C. Gu, and X.-G. Wen, Phys.\ Rev.\ B
\textbf{84}, 235128 (2011).

\bibitem{FidkowskiKitaev2010}
L.\ Fidkowski and A.\ Kitaev, 
Phys.\ Rev.\ B \textbf{81}, 134509 
(2010).

\bibitem{FidkowskiKitaev2011}
L.\ Fidkowski and A.\ Kitaev, 
Phys.\ Rev.\ B \textbf{83}, 075103
(2011).

\bibitem{Turner2011}
 A. M. Turner, F. Pollmann, and E. Berg, 
 Phys.\ Rev.\ B \textbf{83}, 075102 (2011).

\bibitem{Gu09}
Zheng-Cheng Gu, and Xiao-Gang Wen, Phys.\ Rev.\ B \textbf{80}, 155131 (2009).

\bibitem{Chen2011}
Xie Chen,
Zheng-Cheng Gu,
Zheng-Xin Lin,
and
Xiao-Gang Wen,
Phys.\ Rev.\ B \textbf{87}, 155114 (2013).

\bibitem{Gu12}
Zheng-Cheng Gu, and Xiao-Gang Wen, 
\texttt{arXiv:1201.2648}.

\bibitem{LevinStern2009}
M.\ Levin and A.\ Stern,
Phys.\ Rev.\ Lett \textbf{103}, 196803 (2009).

\bibitem{Neupert2011}
 Titus Neupert, Luiz Santos, Shinsei Ryu,
 Claudio Chamon, and Christopher Mudry,
 Phys.\ Rev.\ B \textbf{84}, 165107 (2011). 

\bibitem{LuVishwanath2012}
Yuan-Ming Lu, Ashvin Vishwanath
Phys.\ Rev.\ B \textbf{86}, 125119 (2012).

\bibitem{LuVishwanath2013}
Yuan-Ming Lu and Ashvin Vishwanath,
\texttt{arXiv:1302.2634}. 

\bibitem{Wang}
J.Wang, and X.-G. Wen, \texttt{arXiv:1212.4863}

\bibitem{Kapustin}
A. Kapustin and N. Saulina, Nucl. Phys. B \textbf{845}, 393 (2011)

\bibitem{Levin 2013}
M. Levin, Phys. Rev. X \textbf{3}, 021009 (2013). 

\bibitem{HanWan2013}
 Ling-Yan Hung and Yidun Wan, Phys.\ Rev.\ B \textbf{87}, 195103(2013), \texttt{arXiv:1302.2951}.

\bibitem{ChengGu2013}
Meng Cheng and Zhen-Cheng Gu,
\texttt{arXiv:1302.4803}. 

\bibitem{LevinGu2012}
M.\ Levin and Z.\ -C.\ Gu, 
Phys. Rev. B \textbf{86}, 115109 (2012). 

\bibitem{LevinGu}
Z.\ -C.\ Gu and M.\ Levin,
\texttt{arXiv:1304.4569}. 

\bibitem{RyuZhang2012}
Shinsei Ryu and Shou-Cheng Zhang,
Phys.\ Rev.\ \textbf{85}, 245132 (2012). 

\bibitem{Laughlin1981}
R. B. Laughlin,
Phys. Rev. B \textbf{23}, 5632 (1981).

\bibitem{orbifolds1}
L. Dixon, D. Friedan, E. Martinec, S. Shenker,
Nucl. Phys. B, \textbf{282,} 13 (1987).

\bibitem{orbifolds2}
L. Dixon, J. Harvey, C. Vafa, E. Witten, Nucl. Phys. B, \textbf{261}, 678(1985).

\bibitem{footnote}
An orbifold projection usually refers to a
symmetry projection which leads to a space-time singularity 
in the string theory context. In this paper, we use the
word orbifold for symmetry projections in general, which
do not necessary generate a space-time singularity. 
More
specfically, the type of projections we discuss in this paper 
include ``shifted orbifolds''
that do not give rise to a
singularity (but  are still called orbifolds). 

\bibitem{assymetric}
K.S. Narain, M.H. Sarmadi and C. Vafa, Nucl. Phys. B \textbf {288}, 551(1987); \textbf{356}, 163 (1991).

\bibitem{LiangQi}
Xiao-Liang Qi, New J. Phys. \textbf{15}, 065002 (2013),
\texttt{arXiv:1202.3983}. 

\bibitem{YaoRyu2012}
Hong Yao and Shinsei Ryu,
\texttt{arXiv:1202.5805}.
  
\bibitem{Cappelli}
A. Cappelli, G. R. Zemba, 
Nucl. Phys. B \textbf{490}, 595 (1997);

A. Cappelli, L. S. Georgiev, G. R. Zemba, 
J. Phys. A \textbf{42} 222001 (2009) ;

A. Cappelli, G. Viola, G. R. Zemba 
Ann.  Phys. (NY) \textbf{325}, 465 (2010);

Andrea Cappelli, Giovanni Viola,
J. Phys. A \textbf{44}, 075401 (2011).

\bibitem{qshe_flux}
Xiao-Liang Qi and Shou-Cheng Zhang, 
Phys.\ Rev.\ Lett.\ \textbf{101}, 086802 (2008).

\bibitem{footnote2}
Orbifolding by a symmetry group is also known as gauging 
since from the point of view of target space physics 
we are using labels in a covering space to label states in the orbifolded theory. Therefore there is a redundancy in the labelling and hence a gauge symmetry.

\bibitem{Cardy1986}
J. L. Cardy, Nucl. Phys. B, \textbf{270}, 186 (1986). 

\bibitem{Polchinski98}
J.\ Polchinski,
\textit{String Theory},
(Cambridge University Press, Cambridge, 1998).

\bibitem{Xiao2013}
X.Chen, O.M. Sule, and S.Ryu (unpublished).

\bibitem{Vafa}
C. Vafa, Nucl. Phys. B \textbf{273}, 592 (1986).
\end{thebibliography}
\end{document}